\documentclass[
  prx,
  twocolumn,
  superscriptaddress,
  showpacs]
  {revtex4-2} 

  \usepackage{graphicx}
  \usepackage{amsmath,amssymb}

\newcommand{\kB}{k_{\text{B}}}

\newcommand{\rmd}{{\rm d}}

\newcommand{\ie}{{\em i.e.}}
\newcommand{\I}{{\em I~}}
\newcommand{\Is}{{\em I}s~}

\newcommand{\pMI}{{\em pMI~}}
\newcommand{\pMIs}{{\em pMI}s~}
\newcommand{\tpMI}{{\it $\tau$-pMI}~}
\def\avg#1{\langle #1\rangle}

\begin{document}

\title{Spontaneous formation of thermodynamically stable Al--Cu--Fe icosahedral quasicrystal from realistic atomistic simulations}

\author{Marek Mihalkovi\v c}
\affiliation{Institute of Physics, Slovak Academy of Sciences, 84511 Bratislava, Slovakia}
\email{mihalkovic@savba.sk}
\author{Michael Widom}
\affiliation{Department of Physics, Carnegie Mellon University, Pittsburgh, PA  15213, USA}
\email{widom@cmu.edu}
\date{\today}

\begin{abstract}
Icosahedral quasicrystals spontaneously form from the melt in simulations of Al--Cu--Fe alloys.  We model the interatomic interactions using oscillating pair potentials tuned to the specific alloy system based on a database of density functional theory (DFT)-derived energies and forces. Favored interatomic separations align with the geometry of icosahedral motifs that overlap to create face-centered icosahedral order on a hierarchy of length scales. Molecular dynamics simulations, supplemented with Monte Carlo steps to swap chemical species, efficiently sample the configuration space of our models, which reach up to 9846 atoms. Exchanging temperatures of independent trajectories (replica exchange) allows us to achieve thermal equilibrium at low temperatures. By optimizing structure and composition we create structures whose DFT energies reach to within $\sim$2 meV/atom of the energies of competing crystal phases.  Free energies obtained by adding contributions due to harmonic and anharmonic vibrations, chemical substitution disorder, phasons, and electronic excitations, show that the quasicrystal becomes stable against competing phases at temperatures above 600K. The average structure can be described succinctly as a cut through atomic surfaces in six-dimensional space that reveal specific patterns of preferred chemical occupancy. Atomic surface regions of mixed chemical occupation demonstrate the proliferation of phason fluctuations, which can be observed in real space through the formation, dissolution and reformation of large scale icosahedral motifs -- a picture that is hidden from diffraction refinements due to averaging over the disorder and consequent loss of information concerning occupancy correlations.
\end{abstract}


\maketitle

\section{Introduction}

Since the discovery of quasicrystals as a distinct phase of matter~\cite{Shechtman}, and recognition of their quasiperiodicity~\cite{LevineSteinhardt}, two fundamental questions remain to be definitively answered: where are the atoms~\cite{Bak}? What stabilizes their quasiperiodic order? Excellent descriptions of their {\it average} structures are possible in terms of cuts through higher dimensional periodic lattices obtained by single-crystal diffraction refinements~\cite{cdyb}. However, quasicrystalline structures can only be reliably equilibrated at high temperatures; consequently, these models contain ambiguous atomic positions with uncertain occupation and chemistry. They omit important {\it correlations} in the case of mixed or partial occupation, and they omit atomic vibrations and diffusion. As regards their thermodynamic stability, local icosahedral motifs are clearly favored energetically, but this need not force long-range quasiperiodicity, as is clearly illustrated by the prevalence of periodic ``approximants'', which mimic quasiperiodic order locally within an ordinary crystalline unit cell that in turn repeats periodically.

An intriguing puzzle that has eluded researchers for three decades is identifying the mechanism that selects an ordered yet non-periodic state. One possible explanation is that quasicrystal are energy minimizing structures, whose structure is forced by specific interatomic interactions~\cite{SocolarMatchingRules}, by maximizing the density of some favorable motif~\cite{Gummelt,Steinhardt}, or by creating a deep pseudogap in the electronic density of states~\cite{Pseudogap,ebands-almnpd,ebands-alconi}. Another possibility is that the structural ambiguity is an {\it intrinsic} characteristic of quasicrystals~\cite{ElserRT,HenleyRT,WidomEntropic}. In this view, the entropy to be gained from chemical or positional fluctuations serves to reduce the free energy relative to competing phases whose energies (without entropy) are lower.  Quasiperiodicity might arise spontaneously because it allows icosahedral symmetry, and this high symmetry maximizes the entropy.

\begin{figure*}
  \includegraphics[width=7.25in]{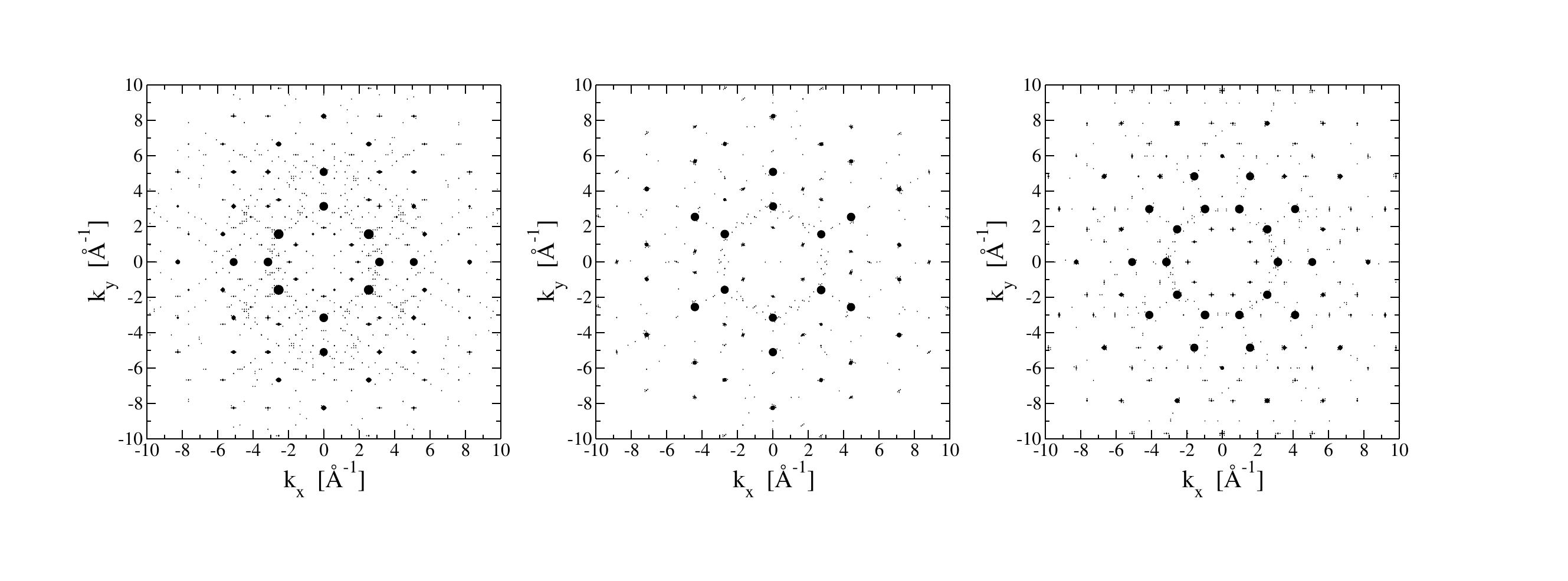}
  \caption{\label{fig:diffract} left to right: 2x, 3x and 5x diffraction patterns of our 9846-atom 8/5 approximant at T=1000K. Diffraction intensities were averaged over 17 independent configurations.}
\end{figure*}

The icosahedral phase of Al--Cu--Fe is an excellent place to seek theoretical insight. Experimentally, the $i$-phase of Al--Cu--Fe was the first well-ordered thermodynamically stable quasicrystal to be discovered~\cite{Tsai_1987}. It exhibits a particular symmetry classified as {\it  face-centered}-icosahedral, and exhibits strong pseudogap in electronic states density near Fermi energy~\cite{pgap-barman}. Recently, these quasicrystals attracted renewed attention following their discovery in a meteorite~\cite{bindi-alcufe}.

In this paper, we report computer simulations leading to spontaneous formation of icosahedral quasicrystals from the melt. Such simulations have been previously reported~\cite{BinaryLJ,Dzugutov,Engel2015}, but only for artificial models that do not describe actual chemical species and hence yield no direct insight into specific experimentally studied compounds.  Here we model Al--Cu--Fe ternary alloys using a combination of chemically accurate density functional theory (DFT) calculations and accurate interatomic pair potentials.  Based on structure models and thermodynamic data provided by our simulation, we calculate a temperature-dependent phase diagram for the Al-rich region of the alloy system showing that energy and entropy conspire in the emergence of the quasicrystal phase as thermodynamically stable at elevated temperatures. Figure~\ref{fig:diffract} illustrates the diffraction patterns of our simulated structures and verifies that we indeed obtain a quasicrystal with the expected face-centered icosahedral symmetry.

\section{Simulation Methods}

Three important ingredients enable the success of our atomistic simulations: realistic DFT-derived interatomic interactions; appropriately sized simulation cells with periodic boundary conditions; efficient hybrid Monte Carlo/molecular dynamics augmented by replica exchange. We exploit the data generated by our simulation to present optimized structure models revealing icosahedral order on a hierarchy of length scales. Combining DFT-calculated formation enthalpies with entropies derived from fluctuations, we calculate the absolute free energies of the quasicrystal phase and competing crystalline phases. Then, from the convex hull of the set of free energies we predict a temperature-dependent phase diagram that shares characteristics with the experimentally assessed behavior, including the emergence of the quasicrystal as a high temperature stable phase.

\begin{figure}[t]
 \includegraphics[clip, width=3.25in]{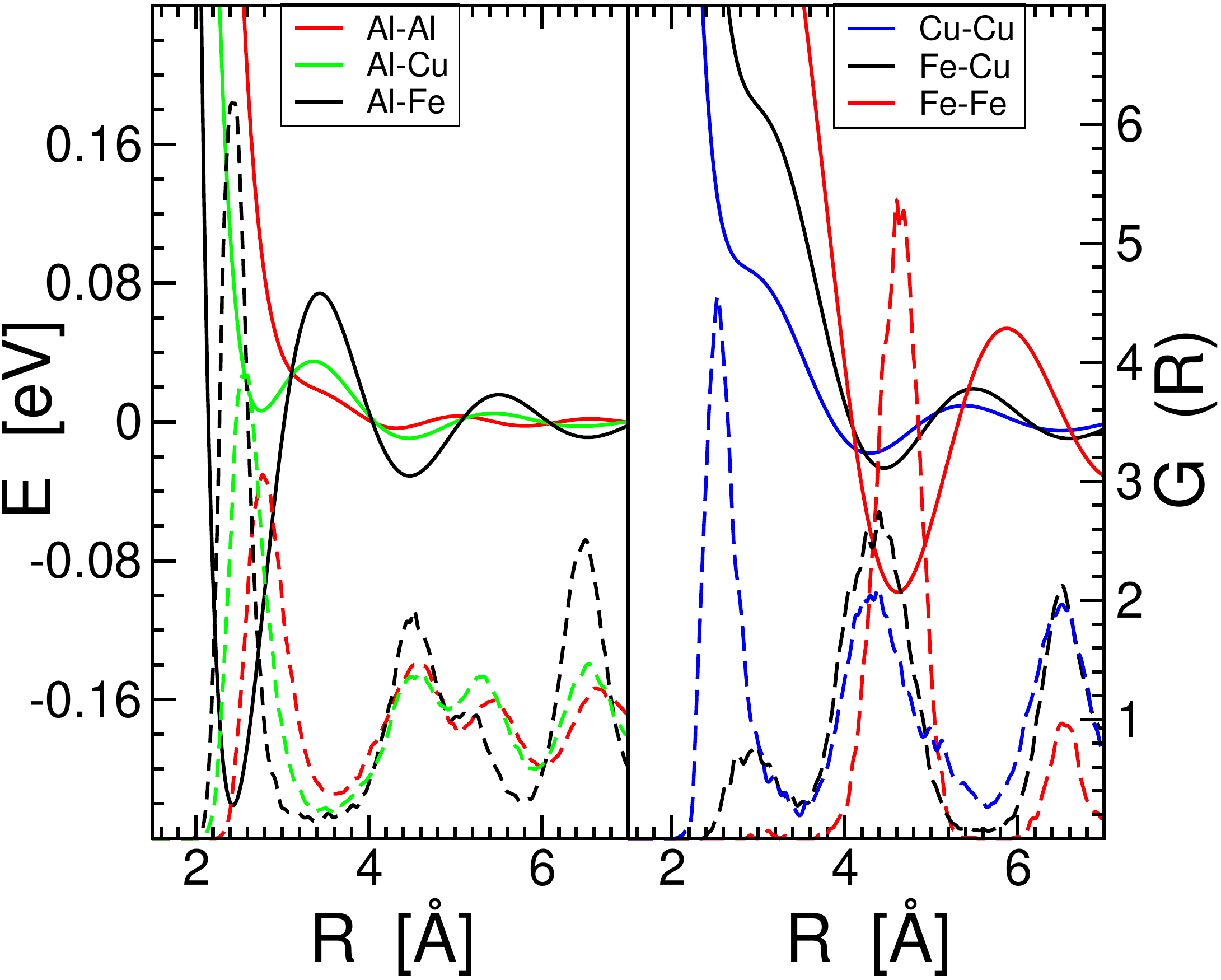}
 \caption{
 \label{fig:ppgofr}
Empirical oscillating pair potentials $V_{\alpha\beta}(r)$ (thick solid lines, energy axis on the left) and partial pair correlation functions $g_{\alpha\beta}(r)$ (dashed lines, vertical axis on the right) for Al--Cu--Fe quasicrystals. The EOPP are fit to a database of DFT energies and forces (see Appendix~\ref{app:eopp}). The pair correlation functions are obtained from single snapshot of the 9846-atom 8/5 approximant EOPP MD simulation.
 }
 \end{figure}

Oscillating interatomic pair potentials accurately describe elemental metals and alloys characterized by weakly-bound $s$ electrons, even in the presence of $s$-$pd$ hybridization.  While these can be derived analytically within electronic density functional theory~\cite{Phillips1994,ico-almn,Moriarty1997}, we instead employ a parametrized empirical form known as EOPP~\cite{EOPP} that we fit to a database of DFT energies and forces (Appendix~\ref{app:dft}). This form has found success modeling many binary and ternary alloys~\cite{mmclh-alir,ScZn,mmclh-mgzndy}.  EOPP also lead to spontaneous formation of single-component icosahedral quasicrystals~\cite{Engel2015}. Henley~\cite{clh-ideas} connected the oscillating potentials with pseudogap near $E_F$ via the Hume-Rothery scenario, and argued that their second minima participate in formation of fundamental clusters. Indeed, under some circumstances the second minima can even create local matching rules favoring quasiperiodicity~\cite{Lim2008}. Figure~\ref{fig:ppgofr} illustrates our fitted potentials and the comparison with interatomic separations in our simulated structures. Details of our fitting procedure are provided in Appendix~\ref{app:eopp}.

 Because the quasicrystal is aperiodic it cannot be precisely represented in a finite size system.  Fortunately, a series of ``rational approximants'' are known that capture the local quasicrystal structure and minimize the deviation caused by application of periodic boundary conditions.
 Reflecting the hierarchical nature of quasiperiodic order, these special sizes grow geometrically as $a_{cub}=2 a_q\tau^n/\sqrt{\tau+2}$ where $\tau=(1+\sqrt{5})/2\approx1.61803...$ is the Golden Mean and $a_q$ is ``quasilattice parameter'' or Penrose Rhombohedron edge length~\cite{eh85}. The volume grows rapidly as $a_{cub}^3\sim \tau^{3n}$. Strictly, face-centered icosahedral symmetry implies $\tau^3$ scaling for self similarity (volume $\sim\tau^9$) along 5--fold and 3--fold directions, but $\tau^1$ scaling along 2--fold icosahedral directions. Here, we label the approximants with a ratio of successive Fibonacci numbers $F_{n+1}/F_n$,  with 128 atoms for ``2/1'' up to 9846 for ``8/5''. Our naming convention is drawn from the definition of Henley's canonical cells designed for packing icosahedral clusters~\cite{clhcct}. Since the fundamental clusters in AlCuFe are $\tau$-times smaller than the proper Mackay clusters decorating the B.C.C. lattice in $\alpha$-AlMnSi~\cite{eh85,elser-model,fujita-double32}, our ``2/1'' approximant has about the same edge length $a_{cub}\sim$12.3\AA~ as $\alpha$-AlMnSi.
 
For small cell sizes (2/1 and 3/2-2/1-2/1), imposing this special length encourages nucleation of the quasicrystal from the melt~\cite{MeltQuench}. In larger cells (3/2, 5/3 and 8/5), the entropic barrier to nucleation is hard to overcome; instead we {\em seed} the structure using the previous approximant size. Because of the discrete cell size scaling, a single unit cell of the seed occupies less than 24\% of the cell volume. Thus we take a supercell of the smaller approximant and enforce the periodic boundary of the bigger approximant. Near the large approximant boundary we let the small approximant overlap itself in a region that is 10\% of the large approximant cell size. This introduces a 25\% excess of atoms in the large approximant placed at unphysically short separations. We remove the excess atoms through a fixed-site lattice gas annealing~\cite{HBSD2} to reach the desired atomic density.

To efficiently anneal both chemical and positional order we utilize a hybrid Monte Carlo/molecular dynamics (MC/MD~\cite{MCMD}) method that samples continuous evolution of atomic positions through molecular dynamics while enabling interchange of chemical species through Monte Carlo swaps. Species swaps are accepted with the Boltzmann probability $\exp{(-\Delta E/\kB T)}$ with $\Delta E$ the change in total energy for the swap. Our simulations are performed in the canonical ensemble with constant temperature, volume and numbers of atoms of each species.  To achieve equilibration at low temperatures and enhance sampling of the configurational ensemble at all temperatures, we supplement our MC/MD simulation with replica exchange~\cite{Swendsen86}.  MC/MD simulations are performed in parallel at many temperatures $T_i$. At fixed intervals we suspend the simulations and consider swapping configurations at adjacent temperatures $T_i$ and $T_{i+1}$. The swap is accepted with a Boltzmann-like probability  based on the energy difference between the configurations. Although the temperature of a configuration jumps during the swap, the configuration remains a properly weighted member of the equilibrium ensemble at its instantaneous temperature.  Further details are in Appendix~\ref{app:replica}.

\section{Results}
\label{sec:results}

We analyze the simulated structures to demonstrate their quasiperiodicity. The clearest demonstration is their diffraction pattern (Fig.~\ref{fig:diffract}) which shows characteristic 2x, 3x and 5x patterns with sharp peaks near the characteristic positions for face-centered icosahedral symmetry (minor deviations occur due to the finite size approximant). The face-centering causes certain diffraction peak positions to occur in ratios of $\tau^3$, while the remainder show $\tau^1$ scaling.

\begin{figure*}[t]
\includegraphics[width=7in]{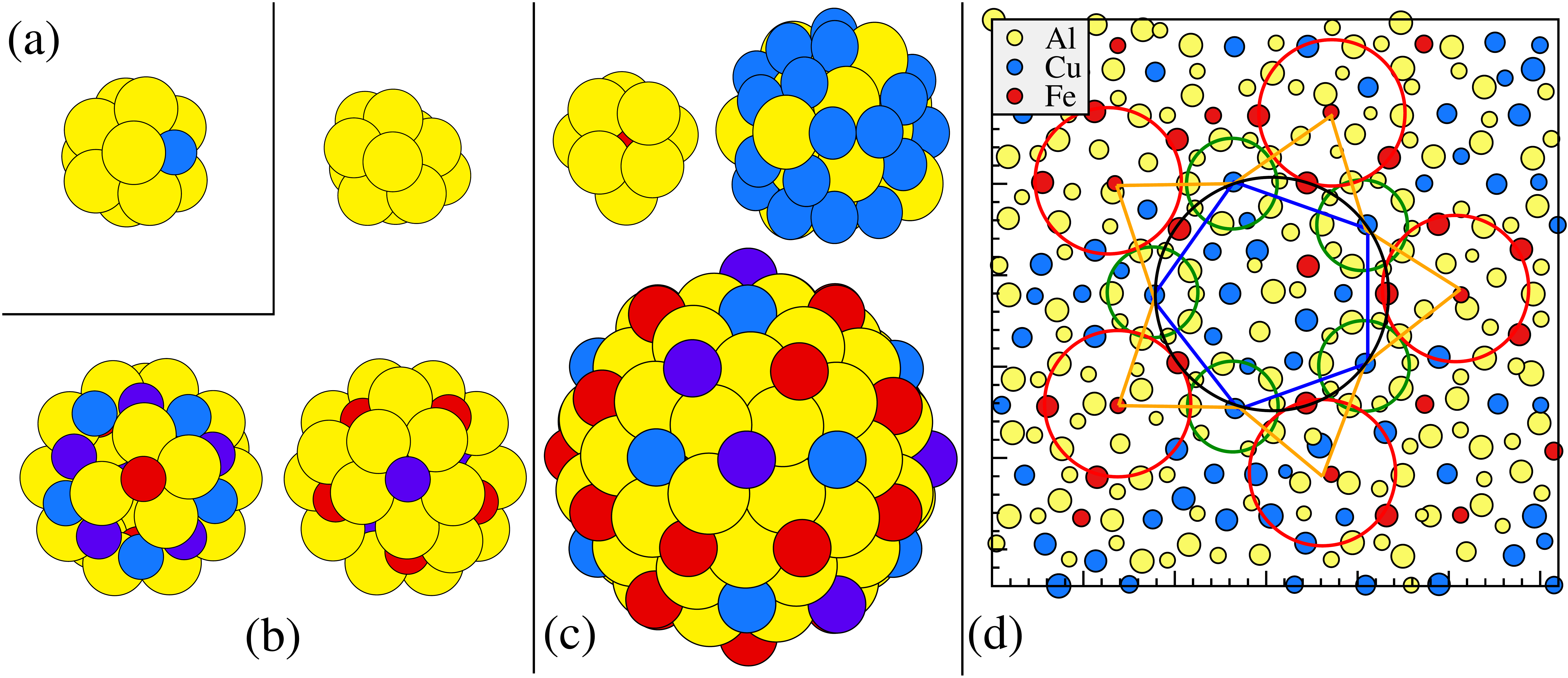}
\caption{\label{fig:clus} Three fundamental clusters constituting the quasicrystal structure. Color coding: yellow (Al), blue (Cu), red (Fe), purple (Cu atoms on mixed Cu/Fe sites).
  (a) Small (Al$_{12-x}$Cu$_x$)Cu icosahedron (\I).
  (b) The pseudo-Mackay icosahedron comprises an inner Al$_{12-x}$Fe shell, an icosahedral (Cu,Fe)$_{12}$ second shell, and an icosidodecahedral (Al,Cu)$_{30}$ third shell. In the example shown here the Cu and Fe atoms in the second shell have arranged to break the icosahedral symmetry down to 5--fold (we show front and back views along the 5--fold axis).
  (c) Four--shell \tpMI cluster with icosahedral Al$_{60}$Cu$_{12}$(Fe,Cu)$_{30}$ outer shell encapsulating an Al$_{12-x}$Fe inner shell and a Cu-rich (Al,Cu)$_{42}$ \pMI-like second and third shell.
  (d) Snapshot of 5/3 approximant from a simulation at T=1200K that we then relaxed. Small icosahedra \I are outlined by green circles, \pMI by red, and an azimuth of the \tpMI by black. Bonds between small icosahedra shown in blue lie along 2--fold directions, bonds between \pMIs shown in red lie along 5--fold directions, and bonds between \Is and \pMIs shown in orange lie along 3--fold directions. Tick marks are placed at 1~\AA~ intervals.
}
\end{figure*}

Examining the structure in real space, we observe that small approximants solidify into well-ordered structures that can be conveniently described as packings of two cluster types. One is a small 13-atom (Al$_{12-x}$Cu$_x$)Cu icosahedron that we denote as \I (see Fig.~\ref{fig:clus}a). The other is a larger pseudo-Mackay icosahedron (\pMI) (Fig.~\ref{fig:clus}b) with a partially occupied Al$_{12-x}$Fe inner shell ($x\sim 1-3$ due to Al-Al repulsion), and a second shell made up of two subshells: a (Cu,Fe)$_{12}$ ``unit-icosahedron'' on the 5--fold axes at ~4.45\AA~ from the center, and an Al-rich (Al,Cu)$_{30}$ icosidodecahedron on the 2--fold axes. In the example shown in Fig.~\ref{fig:clus}b, the Cu and Fe atoms on the unit-icosahedron segregate to break the symmetry from icosahedral to 5--fold. \I-clusters connect along 2--fold icosahedral directions with a spacing of $b=$7.55~\AA, and alternate with \pMI along 3--fold directions ($c=$6.54~\AA~ spacing). This even/odd alternation implements the face-centered icosahedral order.  For the 2/1 approximant this packing is a unique structure producing A, B and C-type canonical cells~\cite{clhcct} (CCT), while the 3/2-2/1-2/1 approximant also contains a symmetry-broken D-cell.

Larger approximants avoid the bulkiest canonical cell D by introducing a new three-shell cluster extending to $\sim$7.7\AA~ (2--fold radius, see Fig.~\ref{fig:clus}c). This cluster is entirely bounded by CCT $Y$-faces, hence it extends, rather than violates, the CCT concept.  Its innermost shell is Al$_{12-x}$Fe as in usual \pMIs, but the second shell, albeit topologically similar to \pMI, is Cu-rich with only ~25\% Al and no Fe atoms. Finally, the third shell is made up from three icosahedral subshells: Al$_{60}$ (at ~6.4\AA), a Fe$_{30-x}$Cu$_x$ ($x\sim$~6) $\tau$-Icosidodecahedron (at 2--fold radius 7.7~\AA) and a Cu$_{12}$ $\tau^2$-Icosahedron (at 5--fold radius 7.2\AA). These 12 Cu atoms are in fact all centers of the small \I clusters; upon including them the whole cluster has 282 atoms.

The three clusters provide a simple, highly economical zero-order description of the structure, since they cover practically all atoms in the structure (99\% in 3/2, 98\% in 5/3 and 97\% in 8/5 approximant). A typical example of these clusters as they appear in our simulations is shown in Fig.~\ref{fig:clus}d, taken from the equilibrium ensemble at T=1200K, followed by relaxation. Mixed chemical occupation breaks the cluster symmetry and can serve as a stabilizing source of entropy, while, together with the cluster covering, it is potentially a means of forcing quasiperiodicity~\cite{Gummelt,Steinhardt}.

\begin{figure}[t]
  \centerline{\includegraphics[width=3.5in]{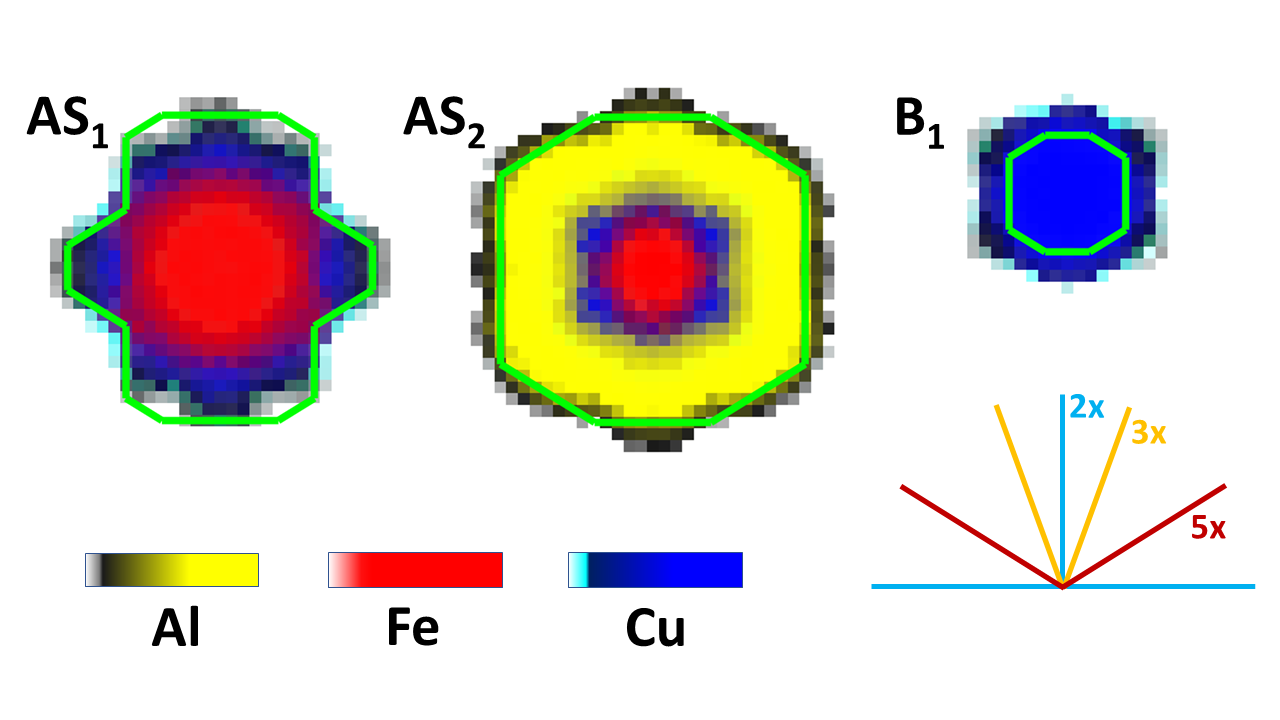}}
  \caption{ 
    \label{fig:perp}
    (top) Atomic surface occupation averaged over ensemble of 3000 configurations from 9846-atom ``8/5'' approximant at T=1242K; (bottom) color bars for chemical species occupancy. Mixed chemical occupation is represented by adding the RGB color values.}
\end{figure}

 The ensemble of structures can be represented using the 6D cut and project scheme. This is an efficient way to represent the average structure in a manner that automatically enforces perfect quasiperiodicity. Details are presented in Appendix~\ref{app:6D}. Approximately 80\% of atoms match projected 6D positions with an accuracy of 0.45~\AA~ or better.  The exceptions are primarily the frustrated Al atoms of the \pMI ~inner shell: their positions are dictated by Al-Al repulsion in the tight inner shell of 9-10 Al atoms around the central Fe, rather than the wells of the Al--TM potential. Three atomic surfaces emerge (see Fig.~\ref{fig:perp}), two large ones (AS$_1$ and AS$_2$) sit at hypercubic lattice sites (``nodes''), one even and the other odd. The remaining small one (B$_1$) sits at the hypercubic body center. Each atomic surface has a particular pattern of chemical occupation. AS$_1$ is primarily Fe, concentrated at the center, with Cu surrounding and finally traces of Al. AS$_2$ is primarily Al, with a small concentration of Fe at the center and Cu separating the Al from the Fe. The remaining surface B$_1$, at the body center, is primarily Cu. The contrast between AS$_1$ and AS$_2$, along with absence of B$_2$, reflects the strong symmetry breaking from simple icosahedral to face-centered lattice.
 
There is a unique connection between the three fundamental clusters (Fig.~\ref{fig:clus}) and the three atomic surfaces. The B$_1$ surface Cu atoms are centers of the \I clusters, the central Cu/Fe part of the AS$_2$ surfaces occupy \pMI clusters, and the Fe core atoms of the AS$_1$ surface are centers of the large \tpMI clusters.

Notice the smooth variations in color (i.e. chemical occupancy). Cu separates Al from Fe on the atomic surfaces while blending continuously into each. Curiously, the location of Cu at the boundary of Fe and Al on the atomic surface is consistent with the position of Cu in the periodic table between transition ($d$-band) metals and nearly free electron ($sp$-band) metals.  Mixed occupation implies swaps in chemical occupation in real space, and low atomic surface densities correspond to partial site occupation. Because these lead to spreading of the occupation domains in perpendicular space, we may identify these chemical species swaps and fractional occupation as types of phason fluctuations.
 
Our simulated atomic surface occupation largely agrees with the popular Katz-Gratias model~\cite{kg}. Specifically, the node vertex surfaces (AS$_1$ and AS$_2$) both transition from Fe at the center, through Cu, to Al around the edges. Only a single body center (B$_1$) is occupied, solely by Cu. Short-distance constraints determined specific shapes of the KG atomic surfaces. In our model these constraints are obeyed through positional correlations so our surfaces have outer shapes that differ from the KG model. Note that the outer shapes are defined by regions of low occupation probability.


After annealing down to low temperatures using the interatomic potentials, we apply DFT to relax the structures to T=0K and compute their enthalpies of formation, $\Delta H$ relative to pure elements, and energetic instabilities, $\Delta E$ relative to the tie-plane of competing crystal structure enthalpies. Table~\ref{tab:ene} summarizes the compositions and formation enthalpies of several competing phases.

\begin{table}
\begin{tabular}{ccccccc}
  name & $\Delta E$ & $\Delta H$ & N$_{at}$ & Al & Cu & Fe \\
       &  meV/at & meV/at & per cell & \% & \% & \% \\
  \hline
  \hline
$\omega$ (tP40) & S  & -280.0 &   40 & 70.0 & 20.0 & 10.0    \\
$\lambda$ (mC102) & S & -361.5 & 102 &72.5 & 3.9 & 23.5 \\
$\alpha$/$\tau_1$ (oC28) & S & -377.4 & 60 & 66.7 & 6.7 & 26.7 \\
$\beta$ (tP16) & S  & -344.3 &   16 & 50.0 & 18.2 & 31.2     \\
$\phi$' (cP50)\footnote{ similar to experimentally known Al$_{10}$Cu$_{10}$Fe $\phi$-phase}      & S  & -267.0 &   50 & 46.0 & 46.0 &  8.0    \\
 \hline
{\it i}-(2/1)          &  +1.8 & -285.4 &  128 & 64.1 & 25.8 & 10.9 \\ 
{\it i}-(3/2)            &  +4.3 & -292.5 &  552 & 63.8 & 23.9 & 12.3 \\
{\it i}-(5/3)            &  +4.0 & -292.6 & 2324 & 65.0 & 22.5 & 12.5\\ 
\hline
\end{tabular}
\caption{\label{tab:ene}Chemistry, instability $\Delta E$ (letter
  ``S'' for stable) and formation enthalpy $\Delta H$ for quasicrystal
  approximants and competing ternary phases at T=0$K$. The two
  variants of the B2--type $\beta$--phase, resulting from annealing
  simulations, are a 2$\times$2$\times$2 Fe-richer supercell with Pearson
  symbol {\it tP16}, and a 3$\times$3$\times$3 supercell with ordered
  vacancies and Pearson symbol {\it cP50}).}
\end{table}

The small approximants exhibit an unusual electronic density of states (eDOS), with a wide and deep pseudogap as is usual in Al-based quasicrystals, and in addition a deeper and very narrow second pseudogap {\it inside} the broad pseudogap (see Appendix~\ref{app:nrg}).  For optimal density and composition, which we achieve by replacing Cu with a combination of Al and Fe, the Fermi level lies inside this second pseudogap. Specifically, we find the effective valence rules of Al=+3, Cu=+1 and Fe=-2 apply, so we can raise or lower $E_F$ by 1 electron without altering the number of atoms through targeted chemical substitutions such as 2Cu$\leftrightarrow$Al$+$Fe. Composition can be shifted at constant effective valence through substitutions such as 3Al+2Fe$\leftrightarrow$5Cu. This rule matches the slope of the quasicrystal and approximant phase fields in the ternary composition space~\cite{alcufe-approximants}. We discovered that neighboring Fe-Fe pairs lead to states in the pseudogap that can be removed by avoiding these pairs. These optimizations can substantially lower the total energy.  However, these structures remain unstable by 2-4 meV/atom relative to competing ordinary crystal phases in the triangle $\eta_2(AlCu)$--$\lambda(Al_3Fe)$--$\omega(Al_7Cu_2Fe)$ as described in Table~\ref{tab:ene}.  Simulated larger approximants, and supercells of the 2/1 approximant, exhibit only the broad pseudogap. Apparently the higher entropy available in larger simulation cells introduces disorder that washes out the detailed structure leading to the narrow second pseudogap, trading a gain in energy for a compensating gain in entropy. It is conceivable that the EOPP potentials are not sufficiently accurate to capture the interactions responsible for the narrow pseudogap feature.

Notice the sequences of enthalpies of formation, $\Delta H$, that decreases monotonically with increasing approximant size. This suggests a possible energetic mechanism favoring quasiperiodicity. However, this is not yet clear, as we have not demonstrated that the energetically optimized structures are more perfect in their quasiperiodicity than representative high temperature structures. Indeed, the decreasing enthalpy is primarily a reflection of increasing Fe content. From the energies relative to the convex hull, $\Delta E$, which are positive and not systematically decreasing, it is likely that any quasicrystal model will be unstable relative to competing crystal phases at low temperatures. Hence, to explain the formation of the quasicrystals from the melt at high temperatures we must seek either a kinetic or a thermodynamic argument.

\begin{figure*}[t]
  \centerline{
    \includegraphics[width=3.4in]{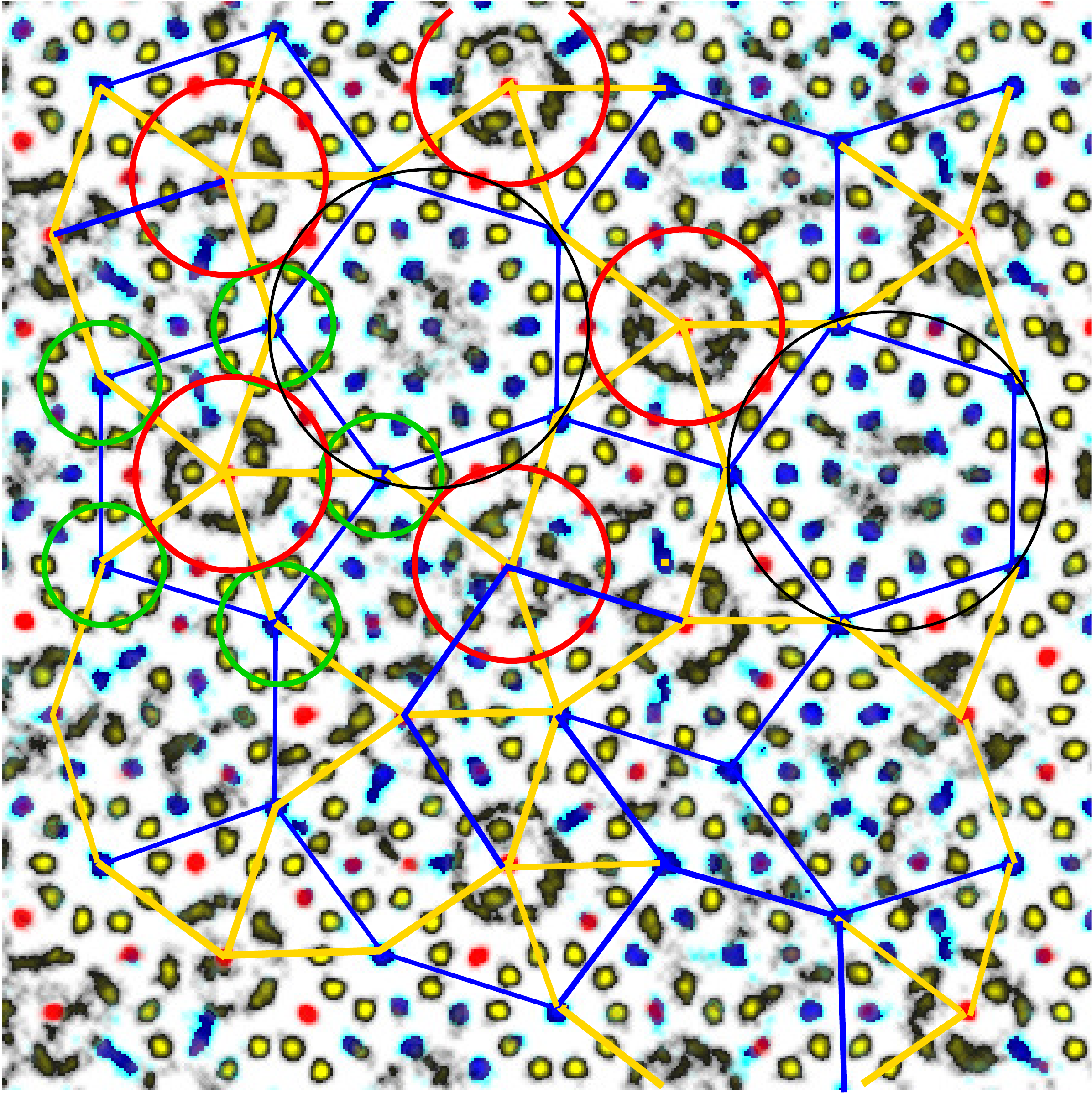}\hspace{0.1cm}
    \includegraphics[width=3.4in]{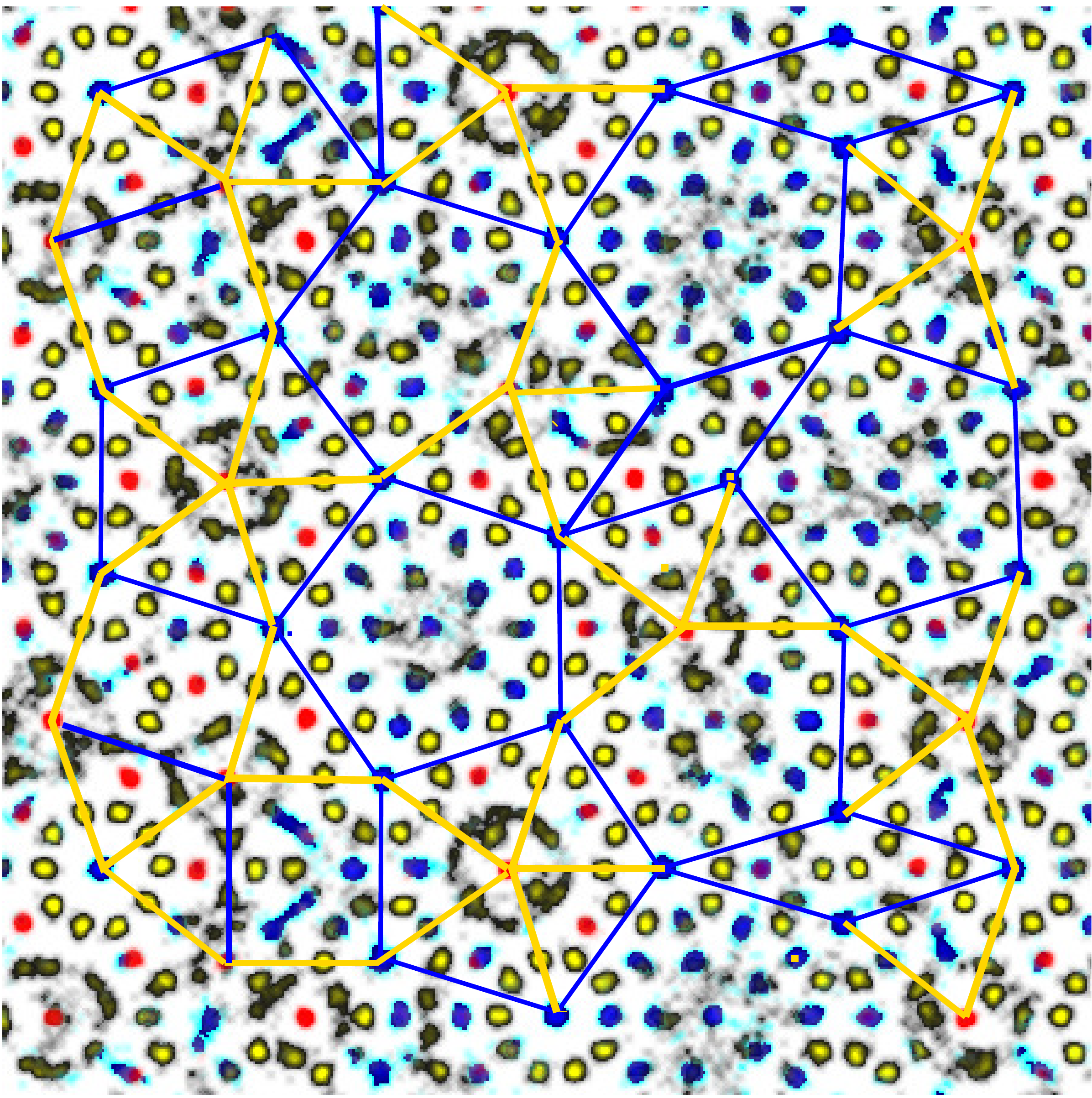}}
  \caption{
    \label{fig:mcmd}
    0.4{\it ns} long exposures from MCMD simulation at 1200K in a
    2304-atom ``5/3'' approximant cubic cell with composition
    Al$_{65.5}$Cu$_{22.0}$Fe$_{12.5}$. The view is parallel to a
    pseudo 5--fold axis, with a slice thickness of 4~{\rm \AA}. {\it
      Left}: Red circles mark 2nd shells of the selected \pMI
    clusters, smaller green circles outline (Al/Cu)$_{12}$Cu
    Icosahedra, and large black circles outline \tpMI clusters.
    Linkages connecting the clusters constitute 2--fold (blue,
    $b=$7.55~\AA) and 3--fold (yellow, $c=$6.54~\AA) linkages of the
    canonical cell tiling. Yellow linkages connect into hexagons,
    boats, stars and decagons analogous to local arrangement familiar
    from decagonal phases. {\it Right}: A new configuration that
    occured 0.8{\it ns} later at the same location.  Defective tilings
    (blue skinny rhombuses) occur at intermediate times.  }
\end{figure*}

We consider the larger approximants, 3/2 (552 atoms) and 5/3 (2324 atoms), as representative of the true quasicrystal. They exhibit interesting structures and dynamics at elevated temperatures.  Because clusters are difficult to identify in individual snapshots due to chemical disorder and atomic displacements, it is best to examine {\it time averages} of the structure. In Fig.~\ref{fig:mcmd} the inner shells of the \pMIs are clearly visible as smeared circles due to the high mobility of the Al atoms, whose positions are frustrated by the incompatible length scale of the icosahedral potential produced by the outer shells with the short-range repulsion of the Al-Al potential. An azimuth of the three-shell \tpMI clusters with Cu-rich interior is indicated by large black circles. Cu-centered small icosahedra are also clearly visible. As time evolves, the identity of these clusters shifts, with some becoming more distinct while others dissolve. Occasionally the structures take pleasing hexagon-boat-star-decagon (HBSD~\cite{HBSD1,HBSD2}) tiling forms as in Fig.~\ref{fig:mcmd}, but these structures, in turn, further evolve. We provide a video illustrating the evolving structure in our Supplemental Online Material.

We find a curious behavior at very high temperatures as the quasicrystals melt. One aspect is that the melting point grows as the approximant size increases. The 2/1 approximant (in 2x2x2 supercell) melts at T=1669 K, the 3/2 at T=1683 K, the 5/3 at T=1723 K, and the 8/5 at T=1788 K. Additionally, the 2/1 and 3/2 melts in a single step, while the 5/3 melts in two(the first broad heat-capacity maximum at 1590 K), and the 8/5 have three additional smaller but sharp heat capacity peaks  at 1266 K, 1471 K and 1680 K that precede melting. Melting of the larger approximants begins with small Al-Cu-rich regions that leave behind an Fe-richer (and hence more mechanically stable) solid quasicrystal phase. Because we perform our simulations at fixed volumes, the actual melting occurs at high pressure - we estimate P=7GPa at the melting point of 8/5. Note that this is consistent with recent experiments of Bindi~\cite{bindi-alcufe-curich}. A video showing liquid-quasicrystal phase coexistence is provided in our Online Supplemental Material. In this video the quasicrystal melts and resolidifies as the liquid interface advances into the solid and then recedes.


Finally, we seek to explain the thermodynamic stability. As shown in Table~\ref{tab:ene} our lowest energy quasicrystal models remain above the convex hull of energy by 2--5 meV/atom.  Thus at low temperatures we anticipate phase separation into the competing phases, $\eta_2(AlCu)$--$\lambda(Al_3Fe)$--$\omega(Al_7Cu_2Fe)$.  Indeed, this is what is shown in the standard phase diagram for the Al--Cu--Fe system~\cite{pdiag-alcufe}.  Finite temperature stability is given by the convex hull of {\it free} energy.  As discussed in Appendix~\ref{app:thermo} the free energy includes harmonic vibrational free energy $F_h$ derived from phonons, electronic free energy $F_e$, and other contributions such as anharmonic phonons, chemical substitution and tiling degrees of freedom, $F_a$.

Comparing free energies of the competing ordinary crystals, small quasicrystalline approximants and the quasicrystal (considered as a large approximant) we predict the stable phases at various temperatures by evaluating the convex hull of free energies over the composition space. Details are presented in Appendices~\ref{app:ternary} and~\ref{app:nrg}. Notably, we observe that the 2/1 approximant gains stability at T=0K when quantum zero point vibrational energy is included, and the 5/3 approximant, which we take as a proxy for the quasicrystal emerges as a stable phase at temperatures above T=600K owing to the excitations contained in $F_a$.

 \begin{figure*}
 \includegraphics[width=3.5in]{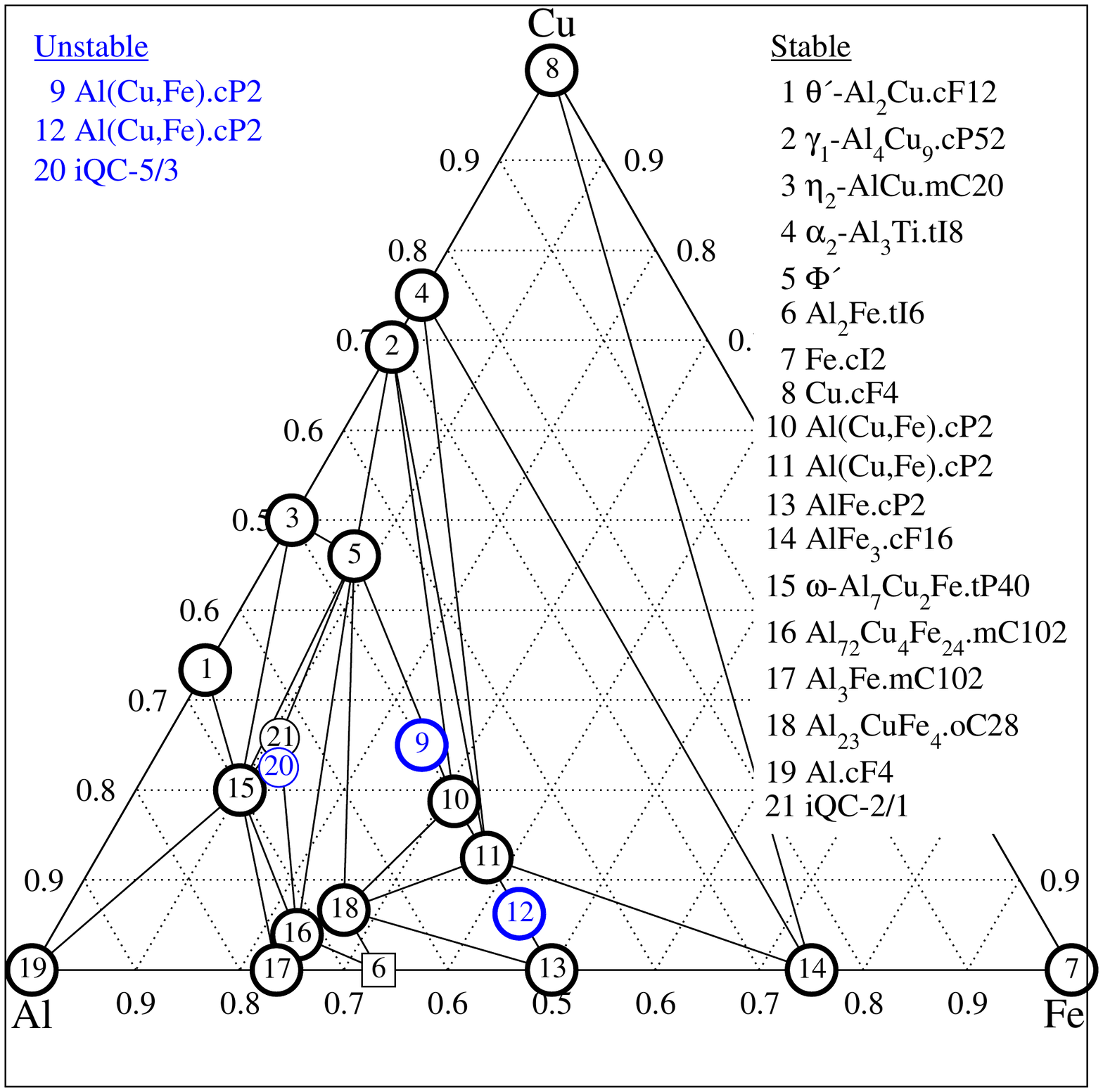}
 \includegraphics[width=3.5in]{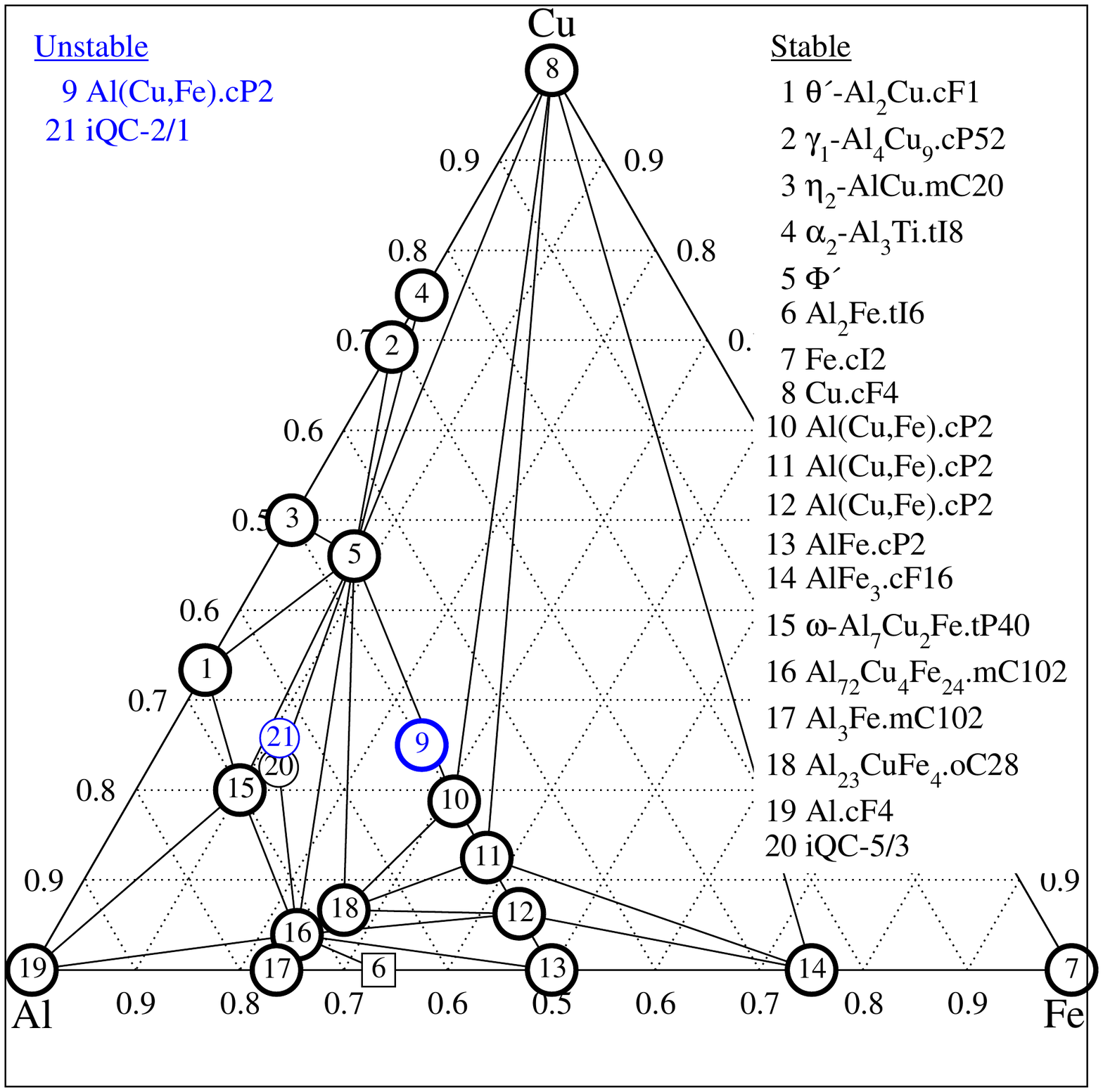}
 \caption{
   \label{fig:pdiag}
   Calculated Al--Cu--Fe phase diagrams at T=0K (left) and T=600K (right). Heavy circles indicate known phases, light circles are quasicrystal approximants, and squares are other hypothetical structures. Black symbols lie on the convex hull while blue lie slightly above, by less than 4 meV/atom. Line segments and enclosed triangles are predicted tie-lines and tie-planes connecting coexisting phases.
}
 \end{figure*}

\section{Discussion}

We have provided four pieces of evidence demonstrating the appearance of a thermodynamically stable quasicrystal state in our model of Al--Cu--Fe. First is the spontaneous formation of quasicrystal approximants, directly from the melt in the case of small approximants, and with the assistance of smaller approximant seeds in the case of large approximants. Second is the appearance of progressively larger clusters and superclusters with increasing approximant size. Third is the decreasing total energy of highly optimized structures with increasing approximant size, suggesting that energy could favor quasiperiodicity. Finally, we have calculated free energies for quasicrystal approximants and competing ordinary crystalline phases and foud that the 5/3 approximant, taken as a proxy for the true quasicrystal, acquires thermodynamic stability above T=600K.

Our findings shed light on the underlying reasons for quasicrystal formation and suggest there is not a single explanation but rather a coincidence of favorable conditions. Although the formation enthalpies decrease (\ie~become more negative, see Table~\ref{tab:ene}) with increasing approximant size, they actually {\em lose} stability relative to competing crystal states owing to the slope of the convex hull with increasing Fe content. The largest cluster for which we have DFT energies, namely 5/3, is the {\em smallest} approximant that can accomodate the \tpMI supercluster including its complete surrounding I clusters. If we postulate that this is an energetically favorable motif then it may be that DFT energies are needed for yet larger approximants before we can claim existence of an energetic preference for the quasicrystal as compared with finite approximants. Further, the appearance of superclusters is a consequence of quasiperiodicity, not a cause. Indeed both matching rule and random tiling models share this feature provided that these are viewed in a time-averaged sense as in the present case. While our explanation for thermodynamic stability at high temperatures is an example of entropic stabilization, the entropy includes phonon anharmonicity in addition to phason-related chemical species swaps and tile flips.

In conclusion, although our explanation for quasicrystal stability is not simple, it illuminates the complex interplay of multiple factors. These include the composition-dependence of competing phase energies, as well as multiple sources of entropy. Chemical preferences for site classes in cluster motifs and on atomic surfaces are favored by our EOPP interatomic potentials. We remark that cluster overlap together with cluster symmetry breaking provides a possible mechanism to force quasiperiodicity, but this appears insufficient to stabilize the Al-Cu-Fe quasicrystal against competing ordinary crystalline phases at low temperature. In our model, the quasicrystal is a high temperature phase.

\begin{acknowledgments}
This work was supported by the Department of Energy under grant DE-SC0014506. Assistance with videos was provided by the Pittsburgh Supercomputer Center under XSEDE grant DMR160149.
M. M. also acknowledges support from Slovak Grant Agency VEGA (No. 2/0082/17), and from APVV (No. 15-0621). Most of the calculations were performed in the Computing Center of the Slovak Academy of Sciences using the supercomputing infrastructure acquired 	under projects ITMS 26230120002 and 26210120002.
\end{acknowledgments}

\appendix

\section{DFT}
\label{app:dft}

First principles calculations within the approximations of electronic density functional theory (DFT) lie at the foundation of our structural and thermodynamic models.  We employ projector augmented wave potentials~\cite{PAW} in the PW91 generalized gradient approximation~\cite{PW91} as implemented in the plane-wave code VASP~\cite{Kresse96}.  Our $k$-point meshes are increased to achieve convergence to better than 1 meV/atom with tetrahedron integration. We employ the default energy cutoffs.  For T=0K enthalpies, all internal coordinates and lattice parameters are fully relaxed. Finite difference methods are applied to calculate interatomic force constants that we need for low temperature vibrational free energies.  Ab-initio molecular dynamics (AIMD) was performed to generate additional energy and force data for fitting interatomic pair potentials. AIMD calculations contained 544 atoms in cubic cell and utilized only a single $k$-point. For accurate cohesive energies, approximant $k$-meshes were converged to 10$^3$ $k$-points/BZ for 128-atom 2/1, 6$^3$ for 552-atom 3/2, and 2$^3$ for 2324-atom 5/3.

\section{Empirical Oscillating Pair Potentials}
\label{app:eopp}

\begin{figure}[t]
 \includegraphics[width=3.25in]{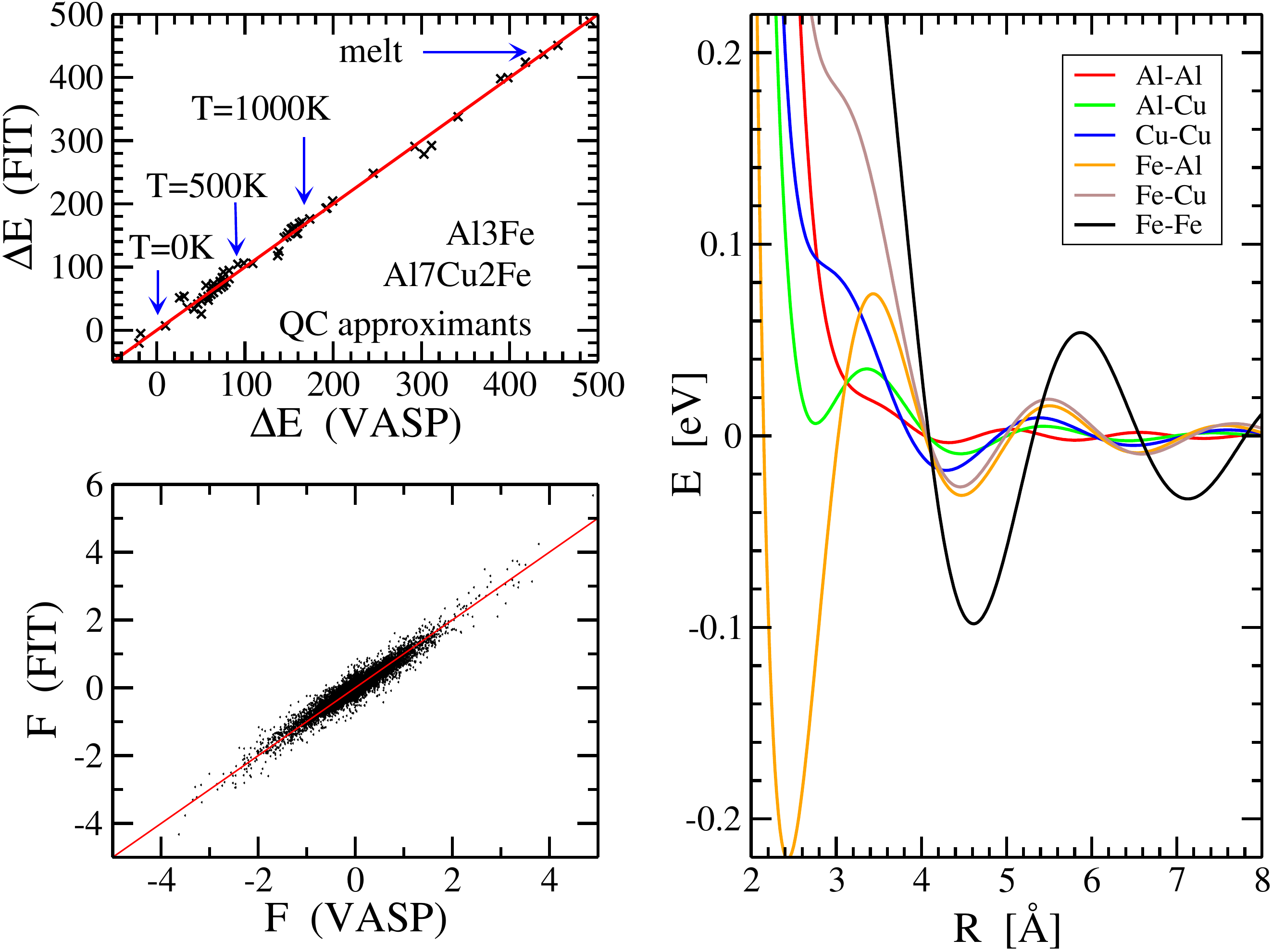}
 \caption{
 \label{fig:ppfit}
EOPP potentials fitted to ab--initio (VASP) force and
 energy data. Parity plots on left are $\Delta E$ (units meV/atom) and $F$ (units eV/\AA). The resulting potentials are shown at right. Parameters are summarized in Table ~\ref{tab:eopp}.
 }
\end{figure}

We choose a 6-parameter empirical oscillating pair potentials (EOPP~\cite{EOPP}) of the form
\begin{equation}
  \label{eq:oscil6}
  V(r) = \frac{C_1}{r^{\eta_1}} + \frac{C_2} {r^{\eta_2}} \cos(k_* r + \phi_*)
\end{equation}
to fit a DFT-derived database of force components and energies. The database contains binary compounds Al$_2$Cu (both tetragonal and cubic), Al$_3$Fe, Al$_5$Fe$_2$, ternary tetragonal Al$_7$Cu$_2$Fe, orthorhombic Al$_{23}$CuFe$_4$, as well as the ternary extension of Al$_3$Fe, and a number of approximant structures. The database contains a significant portion of AIMD data at elevated temperatures, ranging from T=300K up to 2000K, covering both solid and liquid configurations.  In total we use 13176 force-component data points and 63 energy differences (see Fig.~\ref{fig:ppfit}).

\begin{table}
  \begin{tabular}{c|cccccc}
    \hline
    & $C_1$ & $\eta_1$ & $C_2$ & $\eta_2$ & $k_*$ & $\phi_*$ \\
    \hline
Al--Al  &    4337 & 10.416 & -0.1300  & 2.2838  & 4.1702  & 0.8327 \\
Al--Fe  &  1.03$\times$10$^5$ & 17.511 &  4.8643  & 3.3527  & 3.0862  & 1.6611 \\
Al--Cu  &     482 &  8.899 & -2.8297  & 3.7479  & 3.2019  & 4.3551 \\
Fe--Fe  & 1.233$\times$10$^6$ & 13.622 &  5.0695  & 2.5591  & 2.5215  & 3.8725 \\
Fe--Cu  &     461 & 7.363  & -3.7766  & 3.1410  & 2.9191  & 5.7241 \\
Cu--Cu  &    1069 & 9.321  & -2.3005  & 3.2640  & 2.8665  & 0.0586 \\
  \end{tabular}
  \caption{\label{tab:eopp} Fitted parameters for Al--Cu--Fe EOPP potentials.}
\end{table}

We initialized the fit from parameter values that fit GPT potentials~\cite{HBSD2} for a similar system (Al-Co-Ni). The fit quickly converged, with RMS deviation 0.16 eV/\AA~ for forces, and 9.4 meV/atom for energy differences. Since Fe-Fe and Fe-Cu potentials are prone to softening at nearest-neighbor distances due to the lack of data in Al-rich systems, we increased repulsion term coefficients $C_1$ manually for Fe-Fe and Fe-Cu. Final parameters of our potentials are listed in Tab.~\ref{tab:eopp}. Our cutoff radius is taken as 7~\AA.

\begin{figure}[t]
 \includegraphics[width=3.25in]{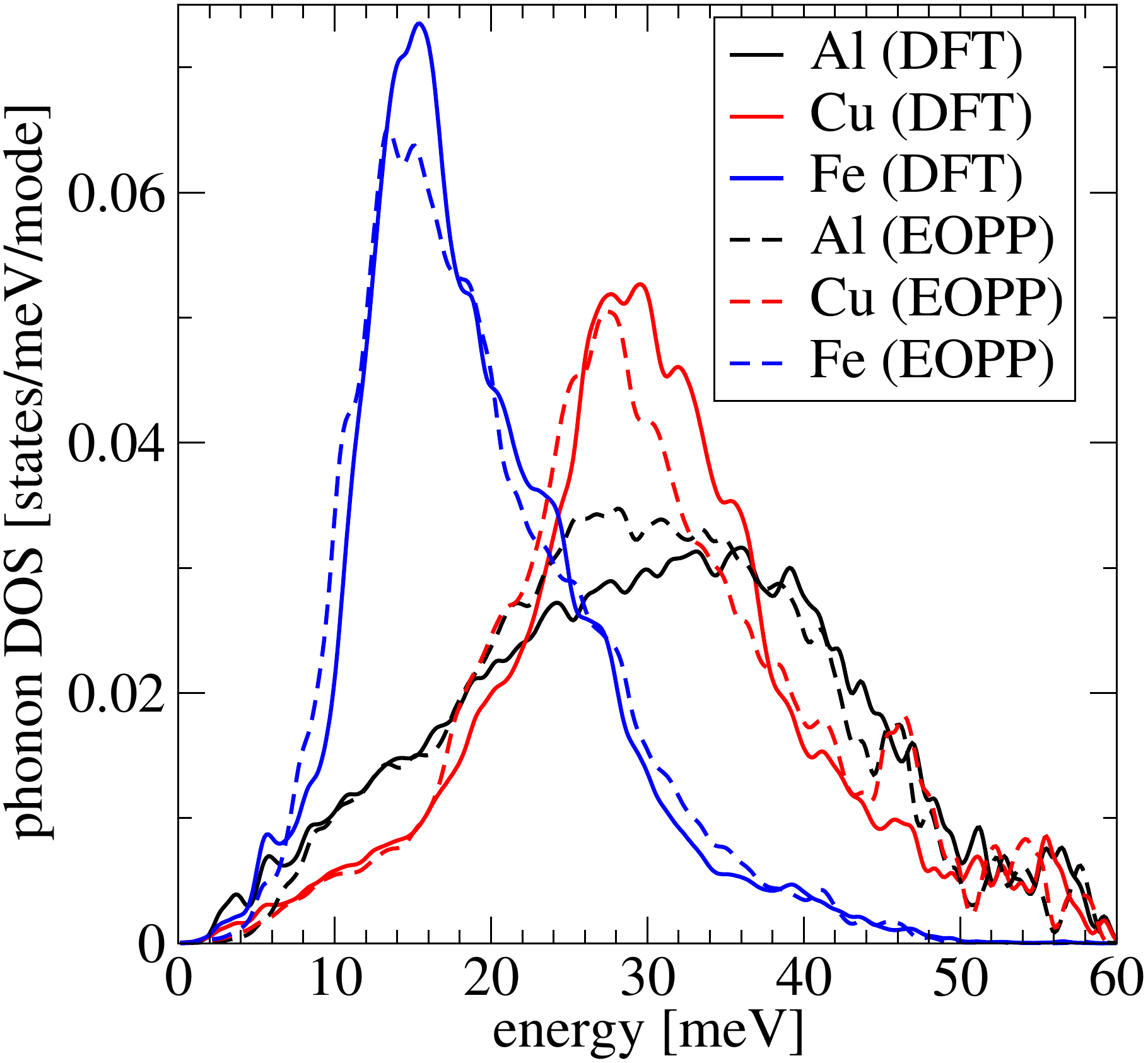}
 \caption{
   \label{fig:pvdos}
   Partial vibrational DOS for a 208-atom orthorhombic approximant of icosahedral quasicrystal, calculated directly by DFT (solid lines), or by EOPP (broken lines).
 }
 \end{figure}

As a demonstration of the accuracy of our pair potentials, we computed the vibrational densities of states (VDOS) for a 208-atom ``3/2-2/1-2/1'' approximant (see Fig.~\ref{fig:pvdos}). The three partials computed by EOPP semiquantitatively match the DFT result, with accuracy comparable to the Sc-Zn case~\cite{ScZn}.

It should be noted that the potentials are valid only for a particular density of the free-electron sea, and they should be used exclusively at constant volume; or in constant--pressure simulations, with additional external pressure set to a value leading to the same equilibrium volume.

\section{Replica Exchange Simulations}
\label{app:replica}

To enhance the efficiency with which our simulations explore the configurational ensemble, we employ a replica exchange mechanism~\cite{Swendsen86}, also known as parallel tempering, in which we perform many runs simultaneously at different temperatures.  The probability for a configuration $i$ of energy $E_i$ to occur at temperature $T_i$ is
\begin{equation}
  P_i =\Omega(E_i) e^{-\beta_i E_i}/Z_i
\end{equation}
where $\Omega(E)$ is the configurational density of states, $Z_i$ is the partition function at temperature $T_i$, and $\beta_i=1/\kB T_i$. The joint probability for configuration $i$ at $T_i$ and configuration $j$ at $T_j$ is
\begin{equation}
  P = P_i P_j = (\Omega(E_i)\Omega(E_j) e^{-(\beta_i E_i+\beta_j E_j)}/Z_iZ_j
\end{equation}
Now consider swapping temperature between configurations $i$ and $j$. The joint probability for configuration $i$ to occur in the equilibrium ensemble at temperature $T_j$ and configuration $j$ to occur at temperature $T_i$ is
\begin{align}
  P' &= (\Omega(E_i)\Omega(E_j) e^{-(\beta_j E_i+\beta_i E_j)}/Z_iZ_j \\
     &= P e^{(\beta_i-\beta_j)E_i-(\beta_i-\beta_j)E_j}
\end{align}
Hence, if the swap is performed with probability $P'/P=e^{\Delta\beta\Delta E}$, equilibrium is preserved following the swap.

This process works most efficiently if energy fluctuations are sufficiently large that the energy distributions $H(E)$ at adjacent temperatures overlap, so that swaps occur frequently.  In this case, a given run (sequence of consecutive configurations) will diffuse between low and high temperatures. Rapid structural evolution at high temperatures thus provides an ongoing source of independent configurations for low temperatures where structural change is intrinsically slow.

We perform isochoric replica-exchange atomistic simulations in the temperature range of 200-1800 K, for several sizes of approximants: 2/1 (128-129 atoms), 3/2 (548 atoms), 5/3 (2324 atoms) and 8/5 (9844 atoms). The basic cycle of the simulation is a species-swap fixed-lattice Metropolis Monte Carlo (MC) stage, consisting of $\sim$30-100 swap attempts per pair, followed by 100-200 MD steps starting directly from the final MC configuration, and finally attempted replica swaps between adjacent temperatures. The temperatures spacing increases linearly with temperature following
\begin{equation}
  \Delta T = \frac{\alpha}{\sqrt{N_{a}}},
\end{equation}
where $\alpha$ is a multiplicative coefficient and $N_a$ number of atoms; such spacing guarantees uniform replica exchange acceptance rates assuming constant heat capacity. Consequently, for large systems an added load is due to the increasingly finer temperature grid. The parameters of most important simulations are summarized in Table~\ref{tab:repex}.

 \begin{table}
   \begin{tabular}{ccccccc}
     L/S &supercell&T-range& $N_a$&MCS& MDS& cycles/10$^3$ \\
     \hline
     8/5 &  1  & 650-1824 & 9844 & 25 & 100 & 33\\
     5/3 &  1  & 200-1800 & 2324 & 30 & 500 & 91\\
     3/2 &$\sqrt{2}\times\sqrt{2}\times 1$  & 400-1800 & 1096 & 25 & 200 & 80\\
     2/1 &$2\times 2\times 2$               & 200-1811 & 1032 & 25 & 200 & 110\\
   \end{tabular}
   \caption{ \label{tab:repex}Cooling simulations for sequence of
     quasicrystal approximants. All sizes have the same composition
     Al$_{65.0}$Cu$_{22.5}$Fe$_{12.5}$. Column MCS is number of Monte
     Carlo attempts per atom for the lattice-gas stage of the
     simulation, MDS number of MD steps (time step $dt$=4fs, per one
     cycle of the simulation. Whole simulation }

 \end{table}

\section{Hyperspace reconstruction}
\label{app:6D}

Lifting a raw atomic structure to hyperspace amounts to associating
the position of each atom to an ideal point in a 6D hypercubic
crystal. We require that the projection of the ideal position into
physical space match the actual atomic position to within a tolerance
$r_{core}$. Additionally we require that the ideal position lie close
in 6D space to the physical 3D space within ``atomic surfaces'' of
definite positions, shapes and sizes. Specifically, we choose
$\tau^2$-triacontahedra at 6D nodes, and unit triacontahedra at 6D
body-centers, of radius
$r_{6D}$=11.7\AA~ in the 5-fold direction. In real space,
the projected sites are separated by at least 1.05\AA. We obtain
unambiguous registration with $r_{core}=$0.45\AA.

We register each structure by: ($i$) identifying all Al$_{12}$Cu
icosahedra in the actual atomic structure (centers of perfectly
icosahedral clusters will have the smallest positional deviation from
ideal sites); ($ii$) mapping these to the even-only {\it body-center}
sublattice (hence resolving the even/odd ambiguity); ($iii$) Given the
relative shift obtained from step ($ii$), we attempt to map the entire
set of atoms positions to the ideal 6D sites. In practice about 80\%
of sites map, with the exceptions being primarily the disordered inner
shells of the \pMI.

Working with periodically bounded boxes in real space results in
finite resolution of the perp-space, namely $d_{res}^\perp =
a_q/\sqrt{F_{n}^2+F_{n-1}^2}$, $F_n$ are Fibonacci numbers and
$a_q$=4.462~\AA. For our 8/5 approximant, $d_{res}^\perp =
a_q/\sqrt{5^2+3^2}\sim$0.765~\AA~.

The atomic surfaces that result from the registration for our largest
8/5 approximant, after averaging over 3000 configurations, are shown
in Fig.~\ref{fig:perp}.

\section{Ternary phase diagrams}
\label{app:ternary}

We model phase stability by exploring the full composition space. In
addition to the quasicrystal phase, we include the pure species in
their favored structures, all known binary Al--Cu and Al--Fe phases,
and all known ternary phases: $\lambda$-Al$_3$Fe.mC102,
$\beta$-AlCuFe, Al$_6$Mn structure type
$\tau_1$-Al$_{23}$CuFe$_4$.oC28, $\omega$-Al$_7$Cu$_2$Fe,
$\phi$-Al$_{10}$Cu$_{10}$Fe. All of the phases have known structures
with the exception of $\beta$ and $\phi$, exhibiting vacancy and
chemical disorder; the $\phi$ phase was claimed to be a superstructure
of Al$_3$Ni$_2$ structure~\cite{zhang} but belongs to the same $B2$--type
family.

To represent the $\beta$ phase family, we evaluated DFT cohesive energies
for every member of the 2x2x2-supercell ensemble of the cubic $B2$
structure type, under the constraint of fixing the cube-vertex occupation as Al.
For a given Cu content, we created a list
of all symmetry-independent Cu/Fe orderings on the {\it
  body--center} sublattice. Ground states with 1-5 Cu atoms per
16-atom supercell revealed that 2-4 Cu atom range yields
stable structures, while 1/16 or 5/16 Cu atom compositions (with 7/16
and 3/16 Fe atom content respectively) are unstable. By examining the
electronic DOS we concluded that the ternary $\beta$ phase is
electronically stabilized at low T by pushing $E_F$ beyond the steep
Fe-$d$-band shoulder, optimally by substituting 3Fe$\rightarrow$3Cu
(per 16-atoms). This ternary $\beta$-phase is predicted to be stable
at T=0K in the composition range 12.5-25\% of Cu.

At the Cu-rich composition, the $\beta$ phase takes a vacancy-ordered
form with experimental composition Al$_{10}$Cu$_{10}$Fe. Since we
could not find any promising T=0K structure in the 2x2x2 supercell, and
larger supercells ensembles are inaccessible to our direct DFT method,
we proceeded with EOPP potentials and fixed-site lattice-gas
annealing~\cite{HBSD2}, in which atoms are constrained to occupy fixed
lattice of sites, but pairs of species with different chemistry are
allowed to swap their positions.  Using this method we discovered a T=0K stable
state in a 3x3x3 supercell, whose composition can be described by a single
parameter $x$=4/(2$\times$3$^2$)$\sim$0.074 and composition
(Al$_{0.5-x}$Cu$_x$)(Cu$_{0.5-2x}$Fe$_x$Vac$_x$) where the parentheses
separate cube-vertex/body-center sublattices respectively.

The quasicrystal family of structures is represented by 2/1
and 5/3 approximants. The former turns out to be the most stable T=0K
structure within the family ($\delta E=+1.8$ meV/atom), while the 5/3
model at ($\delta E=+4.0$ meV/atom is our best representation of the
quasicrystal phase.

To predict the phase diagram at finite temperature, T$>$0, we add the
free energy $F_{\rm Tot}$ in Eq.~(\ref{eq:Ftot}) to the DFT-calculated
enthalpy for each structure considered.  Because full phonon
calculations for large QC approximants are prohibitive, we assume that
they all share the same $F_h$ as the 2/1 approximant. We then compute
the convex hull of the set of free energies (see
Fig.~\ref{fig:pdiag}). Vertices of the convex hull are predicted to be
stable. We also include a few structures whose free energies lie
slightly above the convex hull by up to 4 meV/atom. Structures are
labeled using their phase name followed by their Pearson symbol.

All known structures lie on or near the convex hull, both at
T=0K and at T=600K, with the exception of Al$_2$Fe in its observed
structure of Pearson type aP19. Instead we find the hypothetical
Al$_2$Fe structure of Pearson type tI6 that is believed to be more
energetically stable~\cite{MihalkovicWidom2012}. Some binaries extend
into the ternary composition space, for example Al(Cu,Fe).cP2.  The
two quasicrystal approximants, iQC-2/1 and iQC-5/3 (reference numbers
21 and 20 respectively), swap stability between low and high
temperature. This occurs because the smaller 2/1 approximant has an
optimized structure and composition at which a deep pseudogap appears
and the energy reaches the convex hull, whereas the larger 5/3
approximant lacks a unique optimized structure and composition but
instead enjoys a large anharmonic contribution to its entropy. The 2/1
approximant appears on the convex hull at $T=0$ despite having $E>0$
according to DFT because of the quantum zero point vibrational (or
competing phases) that is contained in $F_h$.  The 5/3 approximant is
the largest for which we can reliably obtain its low temperature
enthalpy by DFT, and hence we take it as a proxy for the true
quasicrystal state.

Thus we predict that the 2/1 approximant should be stable at low
temperatures but transform to the icosahedral quasicrystal at elevated
temperatures of 600K and above. Correspondingly, the quasicrystal
loses stability at low temperature and transforms into the 2/1
approximant. Many actual or implicit transformations have been
reported for Al--Cu--Fe
quasicrystals~\cite{Bancel1989,Audier1991,Bancel1993,Menguy1993}. Fine
detail of the phase diagram at 600 C~\cite{alcufe-approximants}
revealed that except for small window of single-phase stable
quasicrystal composition around Al$_{62}$Cu$_{25.5}$Fe$_{12.5}$, the
quasicrystal phase transforms into a rhombohedral approximant ($R$-5/3
in our notation), and this transformation is
reversible~\cite{menguy-R-alcufe}.  The quasicrystal phase field
shrinks with decreasing temperature~\cite{pdiag-alcufe} but remains
finite at T=560C, a temperature above our predicted transformation. At
low temperatures the kinetics becomes slow and the transformation will
be inhibited, so the quasicrystal remains metastable at low
temperatures.

According to the experimental phase diagram, our energy minimizing, electronically optimized composition (Al$_{65}$Cu$_{22.5}$Fe$_{12.5}$) with $E_F$ in the center of the pseudogap lies in a coexistence region of three phases: $\lambda$, $\omega$ and $R$, and the latter should have composition Al$_{63.5}$Cu$_{24}$Fe$_{12.5}$ in agreement with Ref.~\cite{menguy-R-alcufe}. However, the center of the $R$-approximant phase field is around $x_{Cu}=$0.26 and $x_{Fe}=$0.12, not far from the composition of our low-temperature winner, the 2/1 approximant Al$_{63.3}$Cu$_{25.8}$Fe$_{10.9}$, which lies just outside the $R$-phase stability range.

We simulated the $R$-phase structure in a cell of 718 atoms at our optimized composition of Al$_{65}$Cu$_{22.5}$Fe$_{12.5}$. Although the optimal structure places $E_F$ at the center of a pseudogap, the energy remains higher than the cubic 5/3 approximant by $\sim$ 3 meV/atom. Due to kinetic barriers at low temperatures, existence of the 2/1 approximant cannot be ruled out. Adequate evaluation of all competing phases around 800-1000 K would require systematic variation of composition and density of all competing phases.

\section{Energetic optimization of quasicrystal approximants}
\label{app:nrg}

The most direct comparison between approximants is achieved under the
constraint of equal density/composition revealing the impact of
structure alone. Also, these are the conditions under which the EOPP
energies are most meaningful (see Appendix~\ref{app:eopp}). Expressing
atomic density per $b^3$ volume, where $b\sim$12.2~\AA~ is the side of
the 2/1 cubic cell, we chose density 129.51 atoms/\AA$^3$ and
composition Al$_{65}$Cu$_{22.5}$Fe$_{12.5}$. In order to satisfy this
constraint accurately for all approximants, we worked with the 2/1
approximant in a $2\times 2\times 2$ supercell (1032 atoms), the 3/2
approximant in a $\sqrt{2}\times\sqrt{2}\times 1$ supercell with 1096
atoms, and the 5/3 in its unit cell with 2324 atoms. Supercells were
also required in order to counter the size effect when measuring
anharmonic heat capacity, $F_a$. The resulting optimized energies are
presented in Table~\ref{tab:appene}. The sequence of formation
enthalpies, both for EOPP and full DFT calculations, appears to favor
larger approximants that minimize the phason strain and accommodate
larger superclusters.

Since the optimal density and composition could vary between
approximants, we varied these individually for each approximant within
its conventional unit cell, with the results presented previously in
Table~\ref{tab:ene}. Again the enthalpy is found to be a decreasing
function of approximant size, both for EOPP and for DFT. However, the
energy relative to the convex hull is minimized for the smallest
approximant.  This is because the larger approximants favor greater Fe
content, and the strong bonding of Fe (see Fig.~\ref{fig:ppgofr})
causes a strong slope of the convex hull facets in the direction of
increasing Fe.

\begin{figure}[t]
 \includegraphics[width=3.25in]{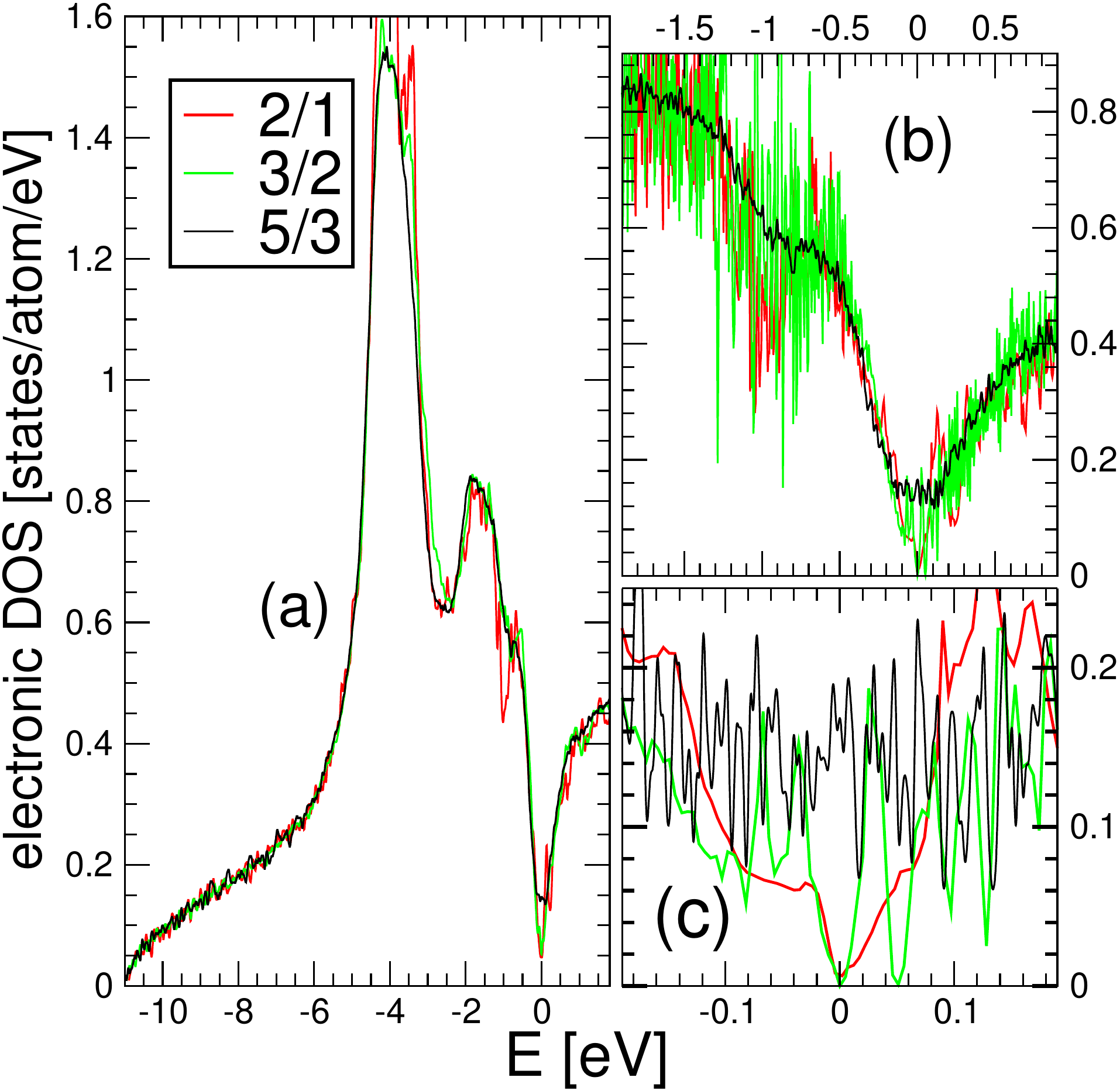}
 \caption{
   \label{fig:edos}
   Electronic density of states for 2/1, 3/2 and 5/3 approximants at their
   optimal compositions (see Table~\ref{tab:ene}), eigenenergies density
   smeared using Gaussian $\sigma$=0.02 eV ($a$). Zoom into
   fine-accuracy tetrahedron method DOS calculation using
   10$\times$10$\times$10 and 6$\times$6$\times$6 meshes respectively
   for 2/1 and 3/2 approximants in ($b$) and ($c$). 5/3 approximant
   (2$\times$2$\times$2 calculation) with resolutions $\sigma$=0.006
   eV ($b$) and 0.002 eV ($c$).  }
 \end{figure}

The 2/1 approximant system size (128-129 atoms) allowed for full
exploration of all degrees of freedom. Under EOPP the icosahedral
structure forms easily from the melt, and we can apply full DFT
refinement to simultaneously explore compositional and density
variation for EOPP pre-optimized models. The structure with lowest
$\Delta E$ occured at increased Cu content, and density of 128
atoms/cell (identical with the Katz-Gratias model
prediction). Starting from this model, we then performed AIMD for
5000 steps (5fs time step) of MD annealing at 1100, 900
and 700 K, and quenched several snapshots from each temperature. The
lowest energy snapshot was from the 900K annealing batch, and yielded the
best atomic structure. This optimal structure exhibits a deep
pseudogap (0.015 states/eV/atom according to tetrahedron method
calculation, see Fig.~\ref{fig:edos}) centered on the Fermi energy.
The pseudogap becomes shallower and broader for structures
in the equlibrium ensemble at higher temperatures.

In the 3/2 approximant cell, we explored several densities: 544 (KG
model density), 552 and 448 atoms/unit cell. The composition was
refined by seeking deepening of the pseudogap and controlling the
Fermi energy assuming a rigid-band picture and simple valence
rules. The best structure was a 552-atom model, greater by 1.5\% than
the KG-model density. This structure was then annealed under AIMD for
5000 steps at 1250K; we observed strong atomic
diffusion and some Al atoms moved as far as 6\AA. Subsequently, we
cooled gradually from 1250K to 800K in another 5000
steps, and finally from 800K down to 300 K in 1000 steps. At
the end, we found that maximal displacement was 5.1\AA~ for Al,
2.9\AA~ for Cu and 0.5\AA~ for Fe atoms; 20 Cu and 100 Al atoms
displaced by more than 0.5\AA. Despite that, the energy of the final
annealed configuration was nearly identical to the initial configuration. Our
best 3/2 approximant has a narrow true gap (according to the
tetrahedron method) at E$_F$.

The 5/3 approximant system size did not allow for systematic
variations, but we did explore several densities (2304, 2324, 2338
atoms) and compositions, using pseudogap depth and Fermi level
position as guides. The final structure was optimized under ab-initio
relaxation in $\sim$70 ionic steps. AIMD annealing was not
feasible. In contrast to the smaller approximants, all 5/3 models
considered had broader and less deep pseudogaps. Nonetheless, the 5/3
approximant achieves the lowest formation enthalpy of the
approximants, most likely as a result of its greatest Fe content.

\begin{table}
  \begin{tabular}{cccccc}
    &$N_{at}$ & $\Delta$E$_{eopp}$ & H$_{DFT}$ & $\Delta$H$_{DFT}$ &  $\Delta$E$_{DFT}$ \\
    \hline
    5/3  &2324 & ref. & -293.0 & ref  & +4.0 \\
    3/2  &1096 & +0.9 & -290.1 & +2.9 & +7.1 (+3.1) \\
    2/1  &1032 & +5.3 & -284.6 & +8.4 &+12.50 (+8.5)\\
    \hline
    8/5 &9846 & +0.3 &    --   & --  \\
   \end{tabular}
   \caption{ \label{tab:appene} Energetic competition (in meV/atom)
     within cubic approximant family at equal composition
     Al$_{65.0}$Cu$_{22.5}$Fe$_{12.5}$, placing $E_F$ exactly at the
     center of the pseudogap (see Fig.~\ref{fig:edos-fixch}).  2/1 and
     3/2 approximants are modeled in supercells of sizes
     $2\times 2\times 2$ and $\sqrt{2}\times\sqrt{2}\times 1$,
     respectively. The largest 8/5 approximant (last row of the Table)
     is inaccessible to DFT evaluation. Notice the enthalpy H$_{DFT}$
     decreases with system size. Column $\Delta$E$_{DFT}$ from convex
     hull evaluation would be equal to the $\Delta$H$_{DFT}$ if the
     approximant compositions were strictly equal. }

 \end{table}

 \begin{figure}[t]
 \includegraphics[width=3.25in]{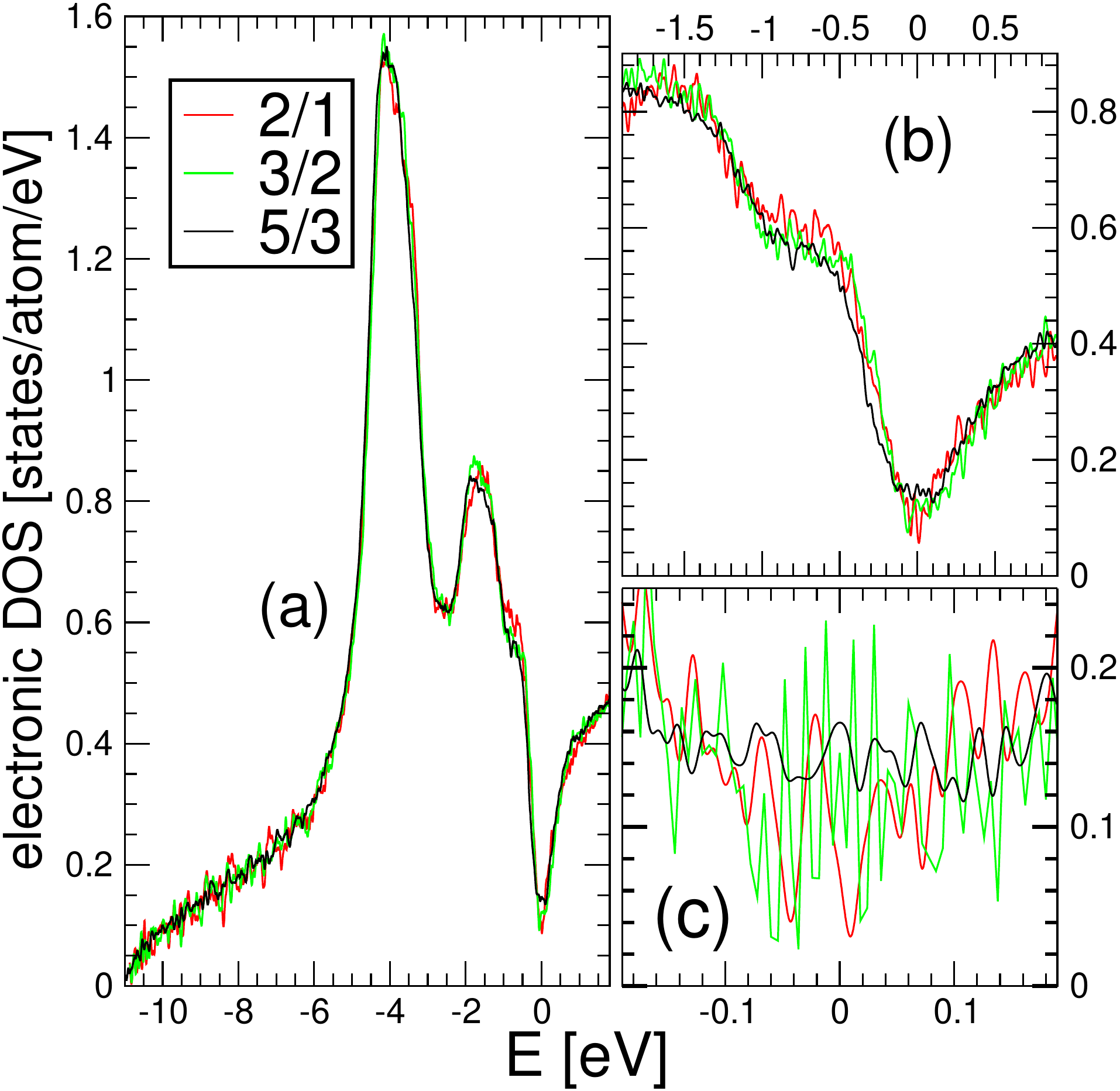}
 \caption{
   \label{fig:edos-fixch}
   Electronic density of states for 2/1, 3/2 and 5/3 approximants at
   fixed composition Al$_{65.0}$Cu$_{22.5}$Fe$_{12.5}$. 2/1 and 3/2
   approximants represented by supercells (see Table~\ref{tab:appene})
   contain amount of frozen disorder comparable to the 5/3
   approximant. Resolution with Gaussian $\sigma$=0.02 eV ($a$), 0.01
   eV ($b$) and 0.006 eV ($c$).  }
 \end{figure}

 \section{Thermodynamics}
 \label{app:thermo}

The Helmholtz free energy $F(N,V,T)$ can be approximately decomposed into a relaxed T=0K energy $E_0$, plus corrections due to harmonic vibrational free energy, $F_h$, anharmonic positional disorder $F_a$, and electronic excitations $F_{\rm elect}$. Thus we write
\begin{equation}
\label{eq:Ftot}
F_{\rm Tot} = E_0+F_h+F_a+F_e.
\end{equation}

The harmonic vibrational free energy of a single phonon mode of frequency $\omega$ is
\begin{equation}
\label{eq:Fmode}
f_h(\omega)=\kB T \ln{[2\sinh{(\hbar\omega/2\kB T)}]}.
\end{equation}
Notice that as $\kB T\ll \hbar\omega$, $f_h(\omega)\rightarrow \hbar\omega/2$, which is the zero point vibrational energy. As $\kB T\gg\hbar\omega$, $f_h(\omega)\rightarrow \kB T\ln{(\hbar\omega/\kB T)}$, which is the classical limit of the free energy. The full harmonic free energy
\begin{equation}
\label{eq:Fh}
F_h(T)=\sum_i f_h(\omega_i).
\end{equation}

The anharmonic contribution includes corrections due to shifts in phonon frequency with large amplitude of oscillation and additional discrete degrees of freedom connected to chemical substitution and possible additional tiling flips. At low temperature these contributions can be neglected, so we only include them beyond a temperature, $T_0$, which we set at 200K. At these elevated temperatures positional degrees of freedom behave nearly classically, so we will evaluate $F_a$ from our classical MC/MD simulations using EOPP.

In the canonical $NVT$ ensemble, the Helmholtz free energy $F(N,V,T)=U-TS$ has the differential
\begin{equation}
\label{eq:dF}
dF(N,V,T)=-S \rmd T -p \rmd V +\mu \rmd N.
\end{equation}
In particular, $S=-\partial F/\partial T$. The entropy is also related to the heat capacity $C=\partial U/\partial T$ through $C/T = \partial S/\partial T$. Hence,
\begin{equation}
\label{eq:C}
C=-T \frac{\partial^2 F}{\partial T^2}.
\end{equation}
Conveniently, $C$ can be obtained at high temperatures from classical MC/MD simulation through fluctuations of the energy,
\begin{equation}
\label{eq:Cfluct}
C=\frac{1}{\kB T^2}\left(\avg{E^2}-\avg{E}^2\right).
\end{equation}
$C$, thus obtained, includes contributions from both harmonic and anharmonic atomic vibrations, and potentially also from chemical substitution and tile flipping.

Because we seek the anharmonic component of the free energy, we define $C_a\equiv C-3N\kB$ for $T>T_0$, and $C_a\equiv 0$ for $T<T_0$. We now integrate $C_a$ once to obtain
\begin{equation}
\label{eq:SofT}
S_a(T)=\int_0^T \frac{C_a(T')}{T'}\rmd T',
\end{equation}
and then integrate once more to obtain
\begin{equation}
\label{eq:Fint}
F_a(T)=-\int_{T_0}^T S_a(T') \rmd T'.
\end{equation}
By definition, $C_a$, $S_a$ and $F_a$ all vanish below $T_0$, then grow continuously at high $T$.

The electronic free energy is obtained from the superposition of single state energies and entropies. An electronic state of energy $E$ is occupied with probability given by the Fermi-Dirac occupation function, $f_\mu(E)=1/(\exp{((E-\mu)/\kB T})+1)$. For $N$ electrons, the chemical potential is defined by the requirement that $N=\sum_i f_\mu(E_i)=N$. Fractional state occupation leads to electronic entropy
\begin{equation}
\label{eq:Se}
S_e(E)=-\kB [f_\mu(E)\ln{f_\mu(E)}+(1-f_\mu(E))\ln{(1-f_\mu(E))}].
\end{equation}
Summing over electronic states, we obtain free energy
\begin{equation}
\label{eq:Fe}
F_e=\sum_i \left(E_i f_\mu(E_i)-T S_e(E_i)\right).
\end{equation}


The overlapping energy distributions at different temperatures created by replica exchange provide an opportunity for accurate calculation of heat capacity, entropy and free energy through the method of histogram analysis. At a single temperature $T$, the frequency distribution of simulated energies $E$ is proportional to the Boltzmann probability
\begin{equation}
  P_T(E)=\Omega(E) e^{-E/\kB T}/Z(T)
\end{equation}
This equation can be inverted to obtain the density of states $\Omega(E)\sim H_T(E) e^{+E/\kB T}$, where $H_T(E)$ is a normalized histogram of energies obtained from a simulation at fixed temperature $T$. Given the density of states, the partition function may be calculated (up to an undetermined multiplicative factor) by integrating,
\begin{equation}
  \label{eq:Zhisto}
  Z=\int{\rm d}E\Omega(E)e^{-E/\kB T}.
\end{equation}
The free energy is determined (up to an additive linear function of T) from
\begin{equation}
  \label{eq:Fhisto}
  F=-\kB T\log{Z},
\end{equation}
and all other thermodynamic functions can be obtained by differentiation. Notice that $Z(T)$ and $F(T)$ are obtained as continuous functions of temperature $T$ over a range of temperatures surrounding the original simulation temperature.

The same approach interpolates between the fixed simulation temperatures by consistently merging densities of states obtained from each temperature~\cite{Swendsen89}. Up to an unknown multiplicative constant, we have
\begin{equation}
  \label{eq:Whisto}
  \Omega(E)=\frac{\sum_T H_T(E)}{\sum_T e^{(F_T-E)/\kB T}}
\end{equation}
where the free energies $F_T$ must be obtained self-consistently with $\Omega(E)$ from Eqs.~(\ref{eq:Whisto}) and~(\ref{eq:Fhisto}).  By setting the value of $F$ and its derivative $S=-\partial F/\partial T$ to values determined from first principles methods at the lowest simulation temperature, we obtain absolute free energy across the entire simulated temperature range.

\bibliography{acf}
 
\end{document}